\documentclass[twocolumn,epjc3]{svjour3}  
\smartqed  
\RequirePackage{graphicx}

  \usepackage{bm}
   \usepackage{amsmath}
    \usepackage{amssymb}
     \usepackage{pifont}
\usepackage[utf8]{inputenc}

\usepackage{xcolor}

\newcommand{\nn}{\nonumber}

\renewcommand{\(}{\left(}
\renewcommand{\)}{\right)}

\journalname{Eur. Phys. J. C}
\begin{document}

\title{What can be learned from the transition form factor $\gamma^*\gamma^*\to\eta'$: feasibility study
}

\titlerunning{Feasibility study for $\gamma^*\gamma^*\to\eta'$ meson transition FF}        

\author{Yao Ji\thanksref{e1,addr1}
        \and
        Alexey Vladimirov\thanksref{e2,addr1} 
}

\thankstext{e1}{e-mail: yao.ji@ur.de}
\thankstext{e2}{e-mail: alexey.vladimirov@physik.uni-regensburg.de}


\institute{Institut f\"ur Theoretische Physik, \\ Universit\"at Regensburg,\\
D-93040 Regensburg, Germany \label{addr1}
}

\date{Received: date / Accepted: date}

\maketitle

\begin{abstract}
We present an analysis of the recent measurement of $\eta'$-meson production by two virtual photons made by BaBar collaboration. It is the first measurement of a transition form factor which is entirely within the kinematic regime of the collinear factorization approach, and thus provides a clean test of QCD factorization theorem for distribution amplitudes (DAs). We demonstrate that the data is in agreement with the perturbative QCD. Also we show that it is sensitive to power corrections to the factorization theorem and to the decay constants. We discuss features of the meson production cross-section and point out the kinematic regions that are sensitive to interesting physics. We also provide estimation of uncertainties on the extraction of DA parameters.
\PACS{PACS code1 \and PACS code2 \and more}
\end{abstract}

\section{Introduction}
\label{intro}

Recently, the measurement of the two-photon-fusion reaction
\begin{eqnarray}\label{process}
e^+(p_a)+e^-(p_b)\to e^+(p_1)+e^-(p_2)+\eta'(p_\eta),
\end{eqnarray}
in the double-tag mode has been reported by BaBar collaboration \cite{BaBar:2018zpn}. This data open the possibility of studying the meson-transition form factor $F(Q_1^2,Q_2^2)$ with both photon virtualities being large, $Q_{1,2}^2\gg \Lambda^2_{\text{QCD}}$. In fact, it is the first measurement of photon-production of meson where QCD factorization theorem could be applied in a truly perturbative regime. Being the opening analysis of this kind, the data \cite{BaBar:2018zpn} have large uncertainties and could not provide any significant restrictions on the models for DAs. However, this is only the first step to a promising future. In this work, we analyze the data \cite{BaBar:2018zpn} within the QCD factorization approach and explore opportunities granted by such double-tag measurements.

On the theory side, the description of form factor with both non-zero virtualities $F(Q_1^2,Q_2^2)$ is essentially simpler in comparison to the description of form factor with a real photon $F(Q^2,0)$. The latter has been measured by several experiments \cite{Berger:1984xk,Aihara:1990nd,Behrend:1990sr,Gronberg:1997fj,BABAR:2011ad}, and has also been the subject of many theoretical studies, see e.g. \cite{Kroll:2011jk,Agaev:2014wna,Chernyak:2014wra}. The simplification comes from the fact that all interaction vertices are within perturbative regime of QCD (whereas, for $F(Q^2,0)$ one must include description for non-perturbative interaction of a quark with the real photon). Therefore, the data \cite{BaBar:2018zpn} provide a clean test of the factorization approach. Our analysis demonstrates an agreement between the measurement and the theory expectations, if one includes higher-twist corrections.

There are several important questions about the meson structure that could be addressed with the help of $F(Q_1^2,Q_2^2)$. The two prominent are: the validity of the state-mixing picture for hard processes, and the size of the gluon component. In this work we demonstrate that the current level of experimental precision is not sufficient to resolve these questions, however, it allows the determination of $\eta-\eta'$ state-mixing constants. In the last part of the paper, we point out the kinematic regions of cross-section that are sensitive to various parameters, and discuss the uncertainty reduction for theory parameters with the increase of data precision.

\section{Theory input}
\label{sec:theory}

The cross-section for the process (\ref{process}) is given by \cite{Budnev:1974de,Poppe:1986dq}
\begin{eqnarray}\label{th:xSec}
\frac{d\sigma}{dQ_1^2dQ_2^2}&=&\frac{\alpha_{em}^4}{2s^2 Q_1^2 Q_2^2} |F(Q_1^2,Q_2^2)|^2 \Phi(s,Q^2_1,Q^2_2),
\end{eqnarray}
where $s=(p_a+p_b)^2$, $(p_{a,b}-p_{1,2})^2=-Q_{1,2}^2$, and $F$ is the $\gamma^*\gamma^*\to\eta'$ transition form factor. The function $\Phi$ accumulates the information about lepton tensor and the phase volume of the interaction. For completeness we present its explicit form in \ref{app:kinematic}. 

In the case of large-momentum transfer, the form factor $F$ can be evaluated within perturbative QCD. In our analysis we consider leading twist contribution and the leading power-suppressed contribution, which originates from twist-3, twist-4 distribution amplitudes (DAs) and meson mass correction. To this accuracy, the form factor reads
\begin{eqnarray}
F=F_{tw-2}+F_{tw-3}+F_{tw-4}+F_{M}+O(Q^{-6})\, ,
\end{eqnarray}
where we omit the arguments $(Q_1^2,Q_2^2,\mu)$ of the form factors for brevity. In the following we provide minimal details on the theory input to our analysis.

\paragraph{Leading twist contribution.} The leading twist contribution has the following form
\begin{eqnarray}
&&F_{tw-2}(Q_1^2,Q_2^2,\mu)=
\\\nn &&~~\sum_{i}C^{i}_{\eta'}(\mu)\int_0^1 dx T^i_H(x,Q_1^2,Q_2^2,\mu)\phi^i_{\eta'}(x,\mu),
\end{eqnarray}
where $i$ is the label that enumerates various $SU(3)$ and flavor channels, $C^{i}_{\eta'}$ are axial-vector couplings (decay constants), $T_H^i$ is the coefficient function, and $\phi_{\eta'}^i$ is the DA for a given channel. 

In our analysis we have considered NLO expression for the leading twist contribution. At this order, one has singlet $(i=\textbf{1})$ and octet $(i=\textbf{8})$ quark channels, and the (singlet) gluon channel $(i=\textbf{g})$. Coefficient functions for the singlet and octet channels are the same $T_H^{\textbf{1}}=T_H^{\textbf{8}}$, and at LO read
\begin{eqnarray}
T_H^{\textbf{1}}(x,Q_1^2,Q_2^2,\mu)&=&T_H^{\textbf{8}}(x,Q_1^2,Q_2^2,\mu)
\\\nn&=&\frac{1}{xQ_1^2+\bar x Q_2^2}+(x\leftrightarrow \bar x)+O(\alpha_s),
\end{eqnarray}
here and in the following we use the shorthand notation $\bar x=1-x$. The NLO expression for quark and gluon ($T^{\textbf{g}}_H$)  coefficient functions have been evaluated in \cite{Braaten:1982yp} and  \cite{Kroll:2002nt}, respectively.

We use the assumption that at the low-energy reference scale $\mu_0=1~$GeV, the singlet and octet DAs coincide, $\phi^{\textbf{1}}(x,\mu_0)=\phi^{\textbf{8}}(x,\mu_0)$. However, generally, singlet and octet DAs are different since they obey different evolution equations. In particular, the singlet DA $\phi^{\textbf{1}}(x)$ mixes with the gluon DA $\phi^{\textbf{g}}(x)$. Therefore, the gluon contribution must also be accounted for, even if the gluon DA is taken to be zero at the reference scale. The evolution equations and anomalous dimensions at NLO can be found in \cite{Dittes:1983dy,Sarmadi:1982yg,Katz:1984gf} (for the collection of formulas see also appendix  B in ref.\cite{Agaev:2014wna}). It is well-known that it is convenient to present DAs as series of Gegenbauer polynomials. The twist-2 quark and gluon DAs for (pseudo)scalar mesons are
\begin{eqnarray}
\label{eq:phi-t2}
\phi_{\eta'}^{\textbf{q}}(x,\mu)=6x\bar x\sum_{n=0,2,..}^\infty a_{n,\eta'}^{q}(\mu) C_n^{3/2}(2x-1),
\\
\phi_{\eta'}^{\textbf{g}}(x,\mu)=30x^2\bar x^2\sum_{n=2,4,..}^\infty a_{n,\eta'}^{g}(\mu) C_n^{5/2}(2x-1).
\end{eqnarray}
In the following we omit the subscript $\eta'$, since it is the only case considered in this work. Coefficients of such expansion do not mix under evolution at LO, however, they do mix at NLO. The leading asymptotic coefficient $a_{0}^{q}=1$ and does not evolve, which corresponds to the electro-magnetic current conservation. Typically, it is assumed that the coefficients of the higher Gegenbauer modes are smaller than the lower ones. In our analysis we include $a_{2,4}^q$ and $a^{g}_{2}$ modes (while we do take into account higher modes during the evolution procedure).

\begin{figure}
\includegraphics[width=0.4\textwidth]{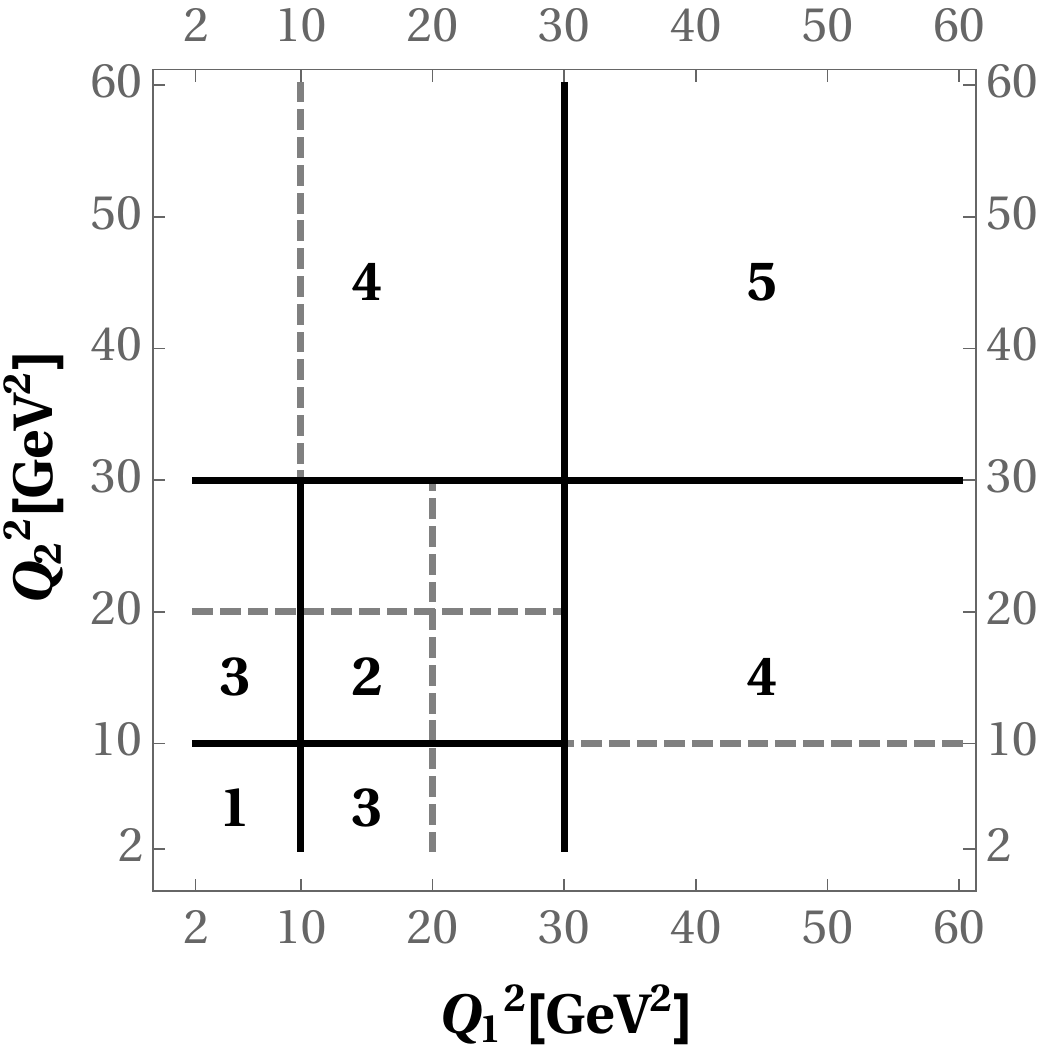}
\caption{The distribution of bins in the plane $(Q_1^2,Q_2^2)$. Black lines and numbers corresponds to bins measured in \cite{BaBar:2018zpn}. Gray dashed lines corresponds to extra binning during the generation of pseudo-data.}
\label{fig:energy-bins}
\end{figure}

\paragraph{FKS scheme.}  We use the Feldmann-Kroll-Stech (FKS) scheme for the definition of couplings $C_{\eta'}^{i}$ \cite{Feldmann:1998vh,Feldmann:1998sh}. The FKS scheme assumes that the $\eta-\eta'$ system can be described as an ideal\footnote{Namely, the coupling constants and wave functions share the same mixing parameters.} mixing of SU(3)-flavor states (singlet and octet). Therefore, the couplings $C_{\eta'}^{(i)}$ can be expressed in terms of quark-couplings with a mixing angle
\begin{eqnarray}
C_{\eta'}^{\textbf{1}}(\mu)&=&C_{\eta'}^{\textbf{g}}(\mu)=\frac{2}{9}(\sqrt{2}f_q \sin \varphi_0+f_s \cos\varphi_0),
\\
C_{\eta'}^{\textbf{8}}&=&\frac{f_q \sin\varphi_0 -\sqrt{2}f_s \cos\varphi_0}{9\sqrt{2}}.
\end{eqnarray}
The values of quark couplings $f_{q}$, $f_s$ and mixing angle $\varphi_0$ are specified later. 

We stress that the coupling  $C^{\textbf{1}}_{\eta'}$ does depend on the scale $\mu$ (whereas, the octet coupling $C^{\textbf{8}}_{\eta'}$ does not). Its dependence appears at NLO due to U(1) anomaly \cite{Kodaira:1979pa} and reads
\begin{eqnarray}
C_{\eta'}^{\textbf{1},\textbf{g}}(\mu)=C_{\eta'}^{\textbf{1},\textbf{g}}(\mu_0)\(1+\frac{2n_f}{\pi \beta_0}(\alpha_s(\mu)-\alpha_s(\mu_0)\),
\end{eqnarray}
where $n_f$ is the number of active flavours. The inclusion of this scale dependence is important for intrinsic consistency of the NLO approximation, but also numerically sizable, e.g the evolution from $1$~GeV to $10$~GeV changes the value of coupling by almost $9\%$.

\begin{figure}
  \includegraphics[width=0.4\textwidth]{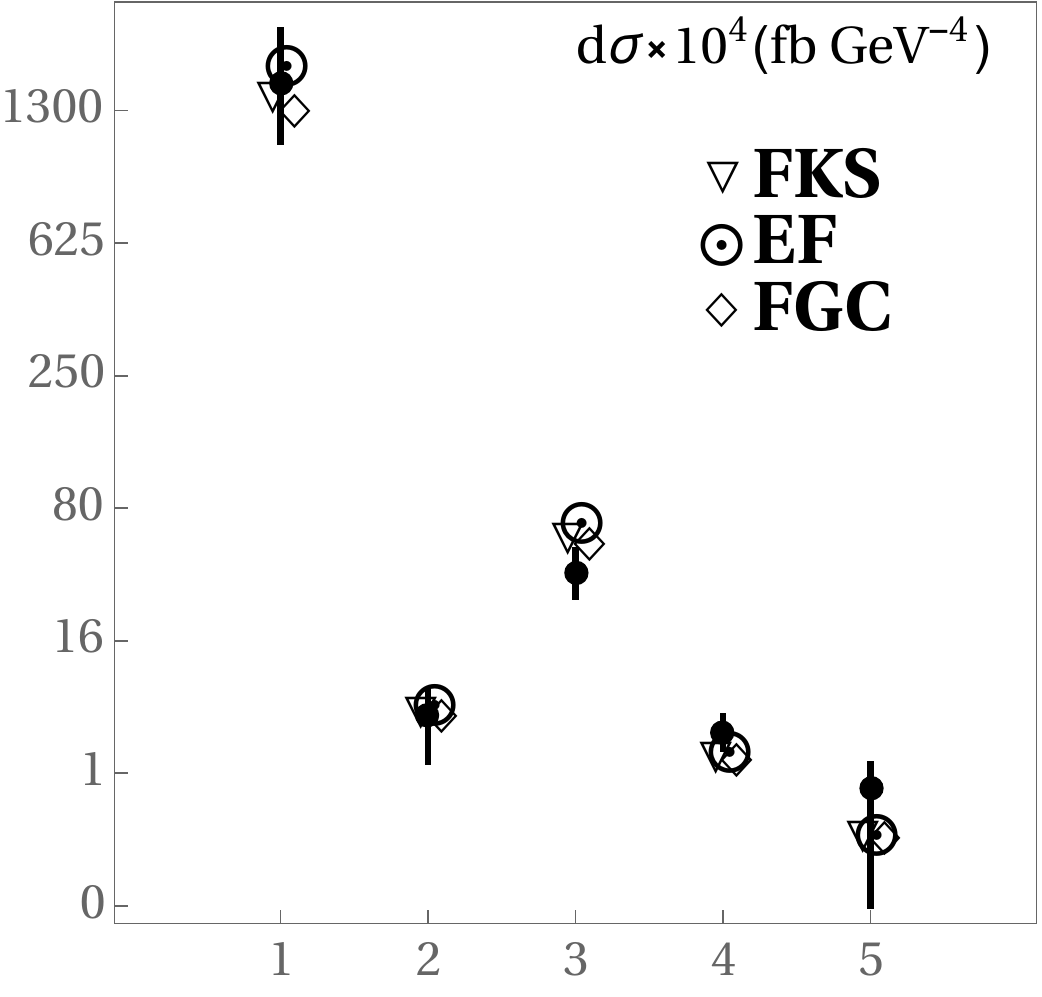}
\caption{Comparison of values of cross-section evaluated in MODEL 1 with different iso-spin coupling parameters to the values of measured cross-section.}
\label{fig:xSec}
\end{figure}
\begin{figure*}
\includegraphics[width=0.33\textwidth]{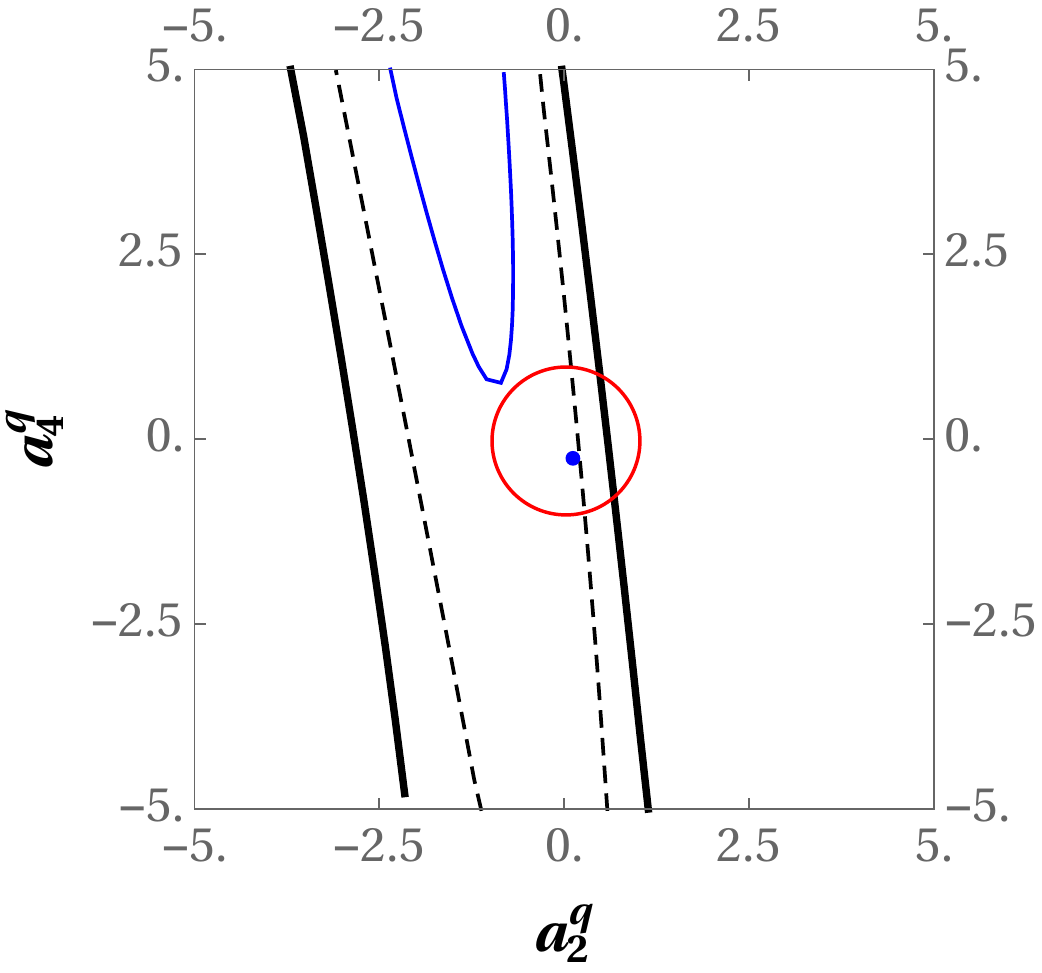}
\includegraphics[width=0.33\textwidth]{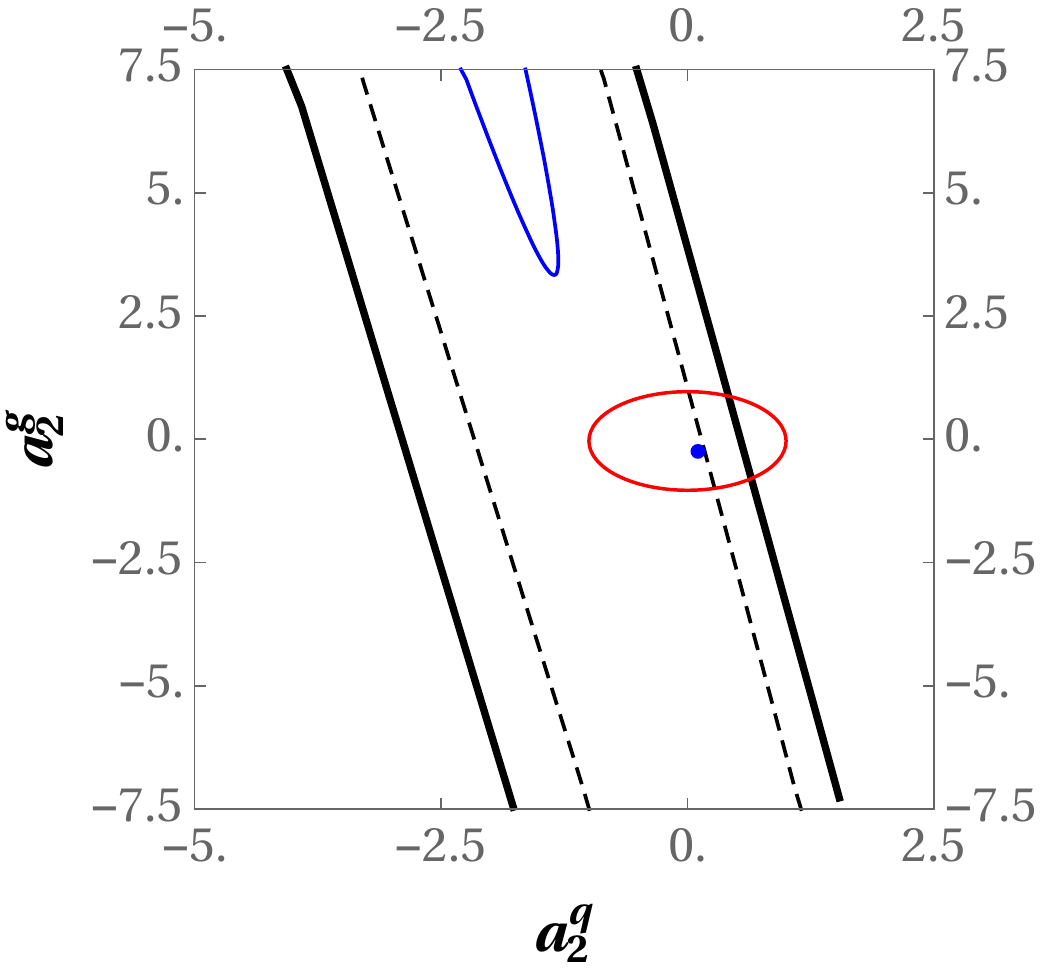}
\includegraphics[width=0.33\textwidth]{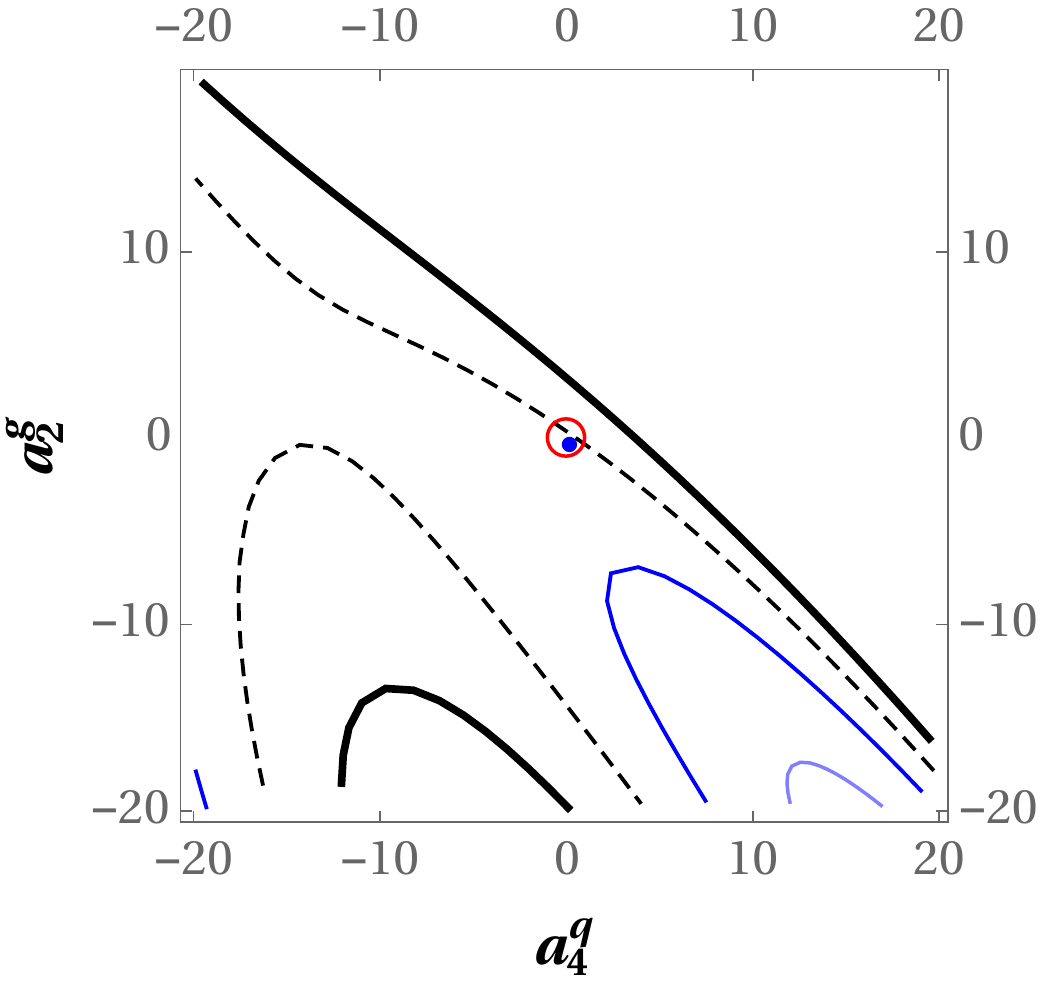}
\caption{The landscape of $\chi^2$ function evaluated for data \cite{BaBar:2018zpn} with FGC parameters in planes of DA moments. The dashed line corresponds to the value $\chi^2/5=1$, the black (blue) corresponds to $\chi^2=6(4)$. The blue dot corresponds to the values of MODEL 1. The red circle designates the approximate region of the theoretical expectation for DA parameters. In each plot, two relevant moments of the DA are varied while the third one is taken from MODEL I.}
\label{fig:landscape}       
\end{figure*}

\paragraph{Target mass correction and higher twist contributions.} As we will demonstrate later, it is important to include the power suppressed contributions in this energy region. These contributions, namely twist-3, twist-4, and the leading meson-mass corrections, have been derived in Ref.~\cite{Agaev:2014wna} in the case of double virtual photons (see also~\cite{Braun:2016tsk}). The expressions of these suppressed contributions have the generic form
\begin{eqnarray}
F_{X}(Q_1^2,Q_2^2,\mu)=\int_0^\infty ds \sum_{q=u+d,s}c_q\frac{\rho^{(q)}_{X,\eta'}(Q_1^2,s,\mu)}{s+Q_2^2},
\end{eqnarray}
where $c_{u+d}=5\sqrt{2}/9$ and $c_s=2/9$. The explicit expressions for the spectral density functions $\rho$ can be found in Ref.~\cite{Agaev:2014wna} as Eq.~(82), (83) and (84) for $\rho_M$,  $\rho_{tw-3}$ and $\rho_{tw-4}$,respectively. The important feature of these corrections is that they all depend on the leading twist Gegenbauer coefficients $a_n$ in Eq.~\eqref{eq:phi-t2}. Importantly, the meson-mass correction does not contain any additional non-perturbative constants, but only parameters from the twist-2 contribution. 

The twist-3 and twist-4 corrections have extra parameters, called $h^{(q)}_{\eta'}$ and $\delta^{(q)}_{\eta'}$. We have used the following values for these constants, determined in \cite{Beneke:2002jn,Bakulev:2002uc},
\begin{eqnarray}\label{th:h}
h_{\eta'}^{(u,d)}&=&0,\qquad h_{\eta'}^{(s)}=(0.5~\text{GeV}^2)\times f_{\eta'}^{(s)}\, ,
\\\label{th:delta}
(\delta_{\eta'}^{(u,d)})^2&=&(\delta_{\eta'}^{(s)})^2=0.2~\text{GeV}^2.
\end{eqnarray}
Strictly speaking, these constants were derived for the case of pion DAs, however, we use these values due to the absence of analogous analysis for $\eta'$. In our study, we have also dropped the quark mass corrections since they only produce a tiny numerical effect.

\section{Analysis of the data}

The measurement \cite{BaBar:2018zpn} provides the differential cross-section $d\sigma/dQ_1^2dQ_2^2$ of $e^+e^-\to e^+e^-\eta^\prime$ measured in five bins. The energy range of bins is shown in Fig.\ref{fig:energy-bins}. The total energy coverage is $2<Q_{1,2}^2<60$ GeV$^2$ that is totally in the range of applicability for the perturbation theory. However, the area of bins is large and thus in order to compare the theory cross-section (\ref{th:xSec}) with the data,  we average the theoretical predictions over each bin. The averaging procedure is essential for such kind of analysis, and could not be replaced by considering the cross-section as a weighted average. This is especially true for the diagonal bins, since the contributions of higher Gegenbauer moments have negligible value at the diagonal $Q_1^2=Q_2^2$.

\paragraph{Input parameters.} The shape of $\eta'$ DA is not very well studied, therefore, there is no commonly accepted values of higher Gegenbauer coefficients. For this initial study we have taken the values discussed in \cite{Agaev:2014wna}. There are three models regarding the leading twist coefficients
\begin{eqnarray}\nn
\text{MODEL 1:}&&
a_2^q=0.10,\quad a_4^q=~~0.1,\quad a_2^g=-0.26,
\\\nn
\text{MODEL 2:}&&
a_2^q=0.20,\quad a_4^q=~~0.0,\quad a_2^g=-0.31,
\\\label{MODELS}
\text{MODEL 3:}&&
a_2^q=0.25,\quad a_4^q=-0.1,\quad a_2^g=-0.22.
\end{eqnarray}
In all these models, the $s$ quark coefficients is assumed to be the same as their $u/d$-quark counterparts. The models are determined at the reference scale $\mu_0=1~$GeV. As for the higher twist corrections, we take the values presented in Eqs.~(\ref{th:h},\ref{th:delta}). In Ref.~\cite{Agaev:2014wna} it was shown that these models are in agreement with the values of the form factor $F(Q^2,0)$ measured by CLEO \cite{Gronberg:1997fj} and BaBar \cite{BABAR:2011ad}.

\begin{table*}
\caption{Values of $\chi^2/\text{\#points}$  evaluated for different theoretical inputs in MODEL 1. The fifth and sixth columns represent values without power corrections (both mass and higher twist) and without higher twist corrections, respectively.}
\label{tab:chi}      
\begin{center}
\begin{tabular}{l||c|c|c||c|c}
\hline\noalign{\smallskip}
 & MODEL 1 & MODEL 2 & MODEL 3 &  No pow. corr. & No tw.3-4 corr. \\
\noalign{\smallskip}\hline\noalign{\smallskip}
FKS & 1.11 & 1.16 & 1.18 & 1.64 & 1.29 \\
EF &  1.83 & 1.92 & 1.98 & 3.12 & 2.29 \\
FGC & 0.97 & 1.00 & 1.02 & 1.35 & 1.08 \\
\noalign{\smallskip}\hline
\end{tabular}
\end{center}
\end{table*}

Other important inputs are the values of the quark couplings $f_{q,s}$ and the $\eta-\eta'$ state-mixing angle $\varphi_0$, defined in the FKS scheme. There are several studies of these parameters. The original work \cite{Feldmann:1998vh} yields
\begin{eqnarray}
\text{FKS}:\qquad
\begin{array}{l}
f_q=(1.07\pm 0.02)f_\pi,\\
f_s=(1.34\pm 0.06)f_\pi,\\
\varphi_0=39.3^o \pm1.0^o.
\end{array}
\end{eqnarray}
Here, and in the following $f_\pi$ is the pion decay constant $f_\pi=103.4\pm0.2$ MeV. The later analysis by Escribano and Freri (EF) \cite{Escribano:2005qq} gives
\begin{eqnarray}
\text{EF}:\qquad
\begin{array}{l}
f_q=(1.09\pm 0.03)f_\pi,\\
f_s=(1.66\pm 0.06)f_\pi,\\
\varphi_0=40.7^o \pm1.4^o.
\end{array}
\end{eqnarray}
Finally, the most recent analysis by Fu-Guang Cao (FGC) \cite{Cao:2012nj} found
\begin{eqnarray}
\text{FGC}:\qquad
\begin{array}{l}
f_q=(1.08\pm 0.04)f_\pi,\\
f_s=(1.25\pm 0.08)f_\pi,\\
\varphi_0=37.7^o \pm0.7^o.
\end{array}
\end{eqnarray}
All these analysis use different data sets and different assumptions, and thus, are competitive to each other.

\paragraph{Test of the theory.} In table \ref{tab:chi}, we show the values of $\chi^2$ per number of points (5 in this case) evaluated within different models. Comparison of values of cross-section (for MODEL 1) is given in Fig.\ref{fig:xSec}. 

One can see from table \ref{tab:chi} that despite the fact that the data is rather poor, it is already rather selective. In particular, the data completely disregards the EF values of iso-spin couplings. It also prefers the FGC values of parameters to the FKS one. Also we see that the data is sensitive to the power corrections, especially to the meson mass correction. We recall that the meson mass correction does not have any new parameters, apart from the state-mixing coupling and DA of the leading twist. The higher twist corrections incorporate parameters $h^q$ and $\delta^q$ in Eqs.~(\ref{th:h},\ref{th:delta}), which in principle, could be extracted from such measurements. 

For FKS and FGC values with power corrections included, we observe perfect agreement of the data with the theory. However, current measurement is not sensitive enough with respect to parameters of DA. All models given in Eq.~(\ref{MODELS}) produce similar results. Moreover, the landscape of the $\chi^2$ function is rather inclusive (see Fig.~\ref{fig:landscape}) and therefore does not allow determination of DA moments. In Fig.~\ref{fig:landscape}, one can see that the parameters $a_n^i$ are strongly correlated in the current data set, and does not even allow accurate determination of the error band. It is a rather unfortunate but predictable conclusion. Indeed, from the five presented bins only two are significantly influenced by the parameters of DAs, as we show in the next section.

\section{Feasibility study}

\begin{figure*}[t]
\includegraphics[width=0.33\textwidth]{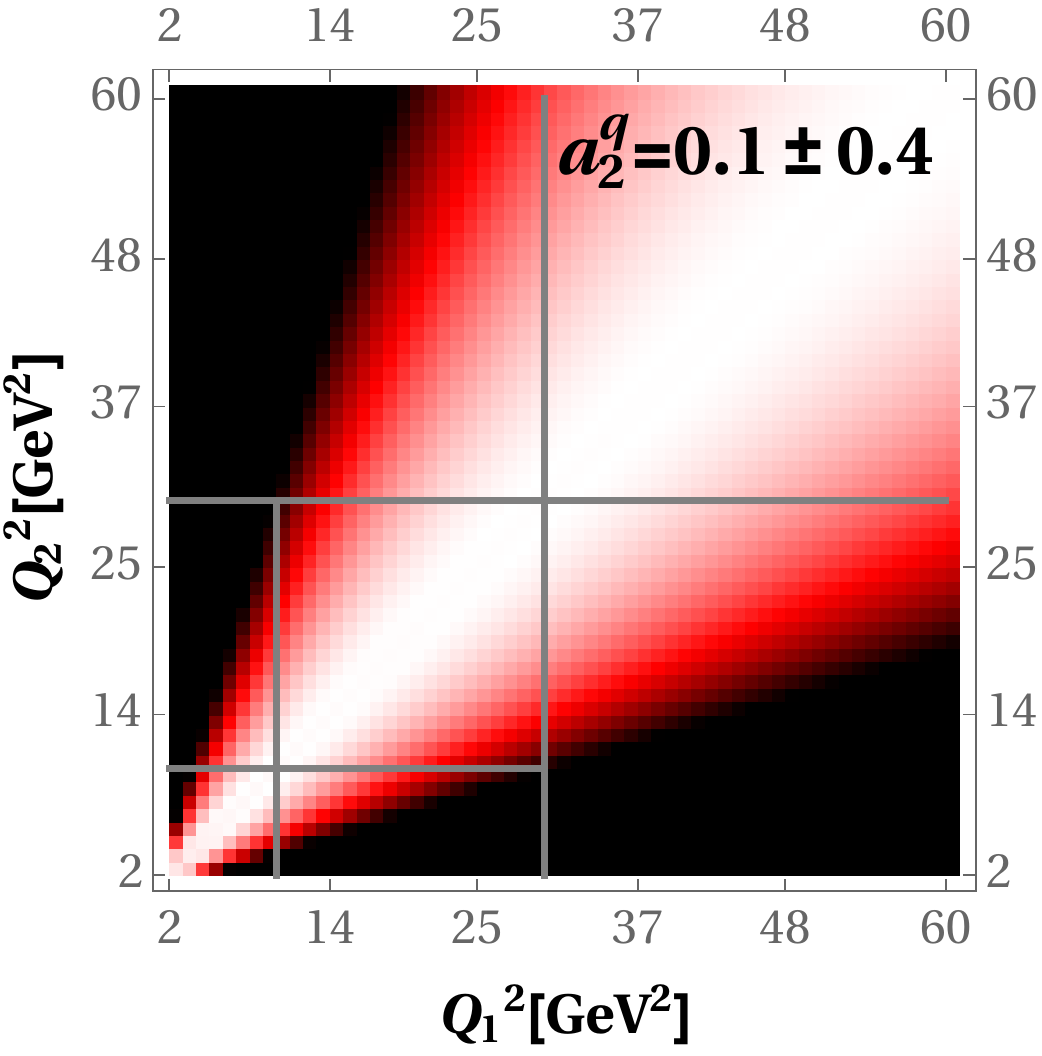}
\includegraphics[width=0.33\textwidth]{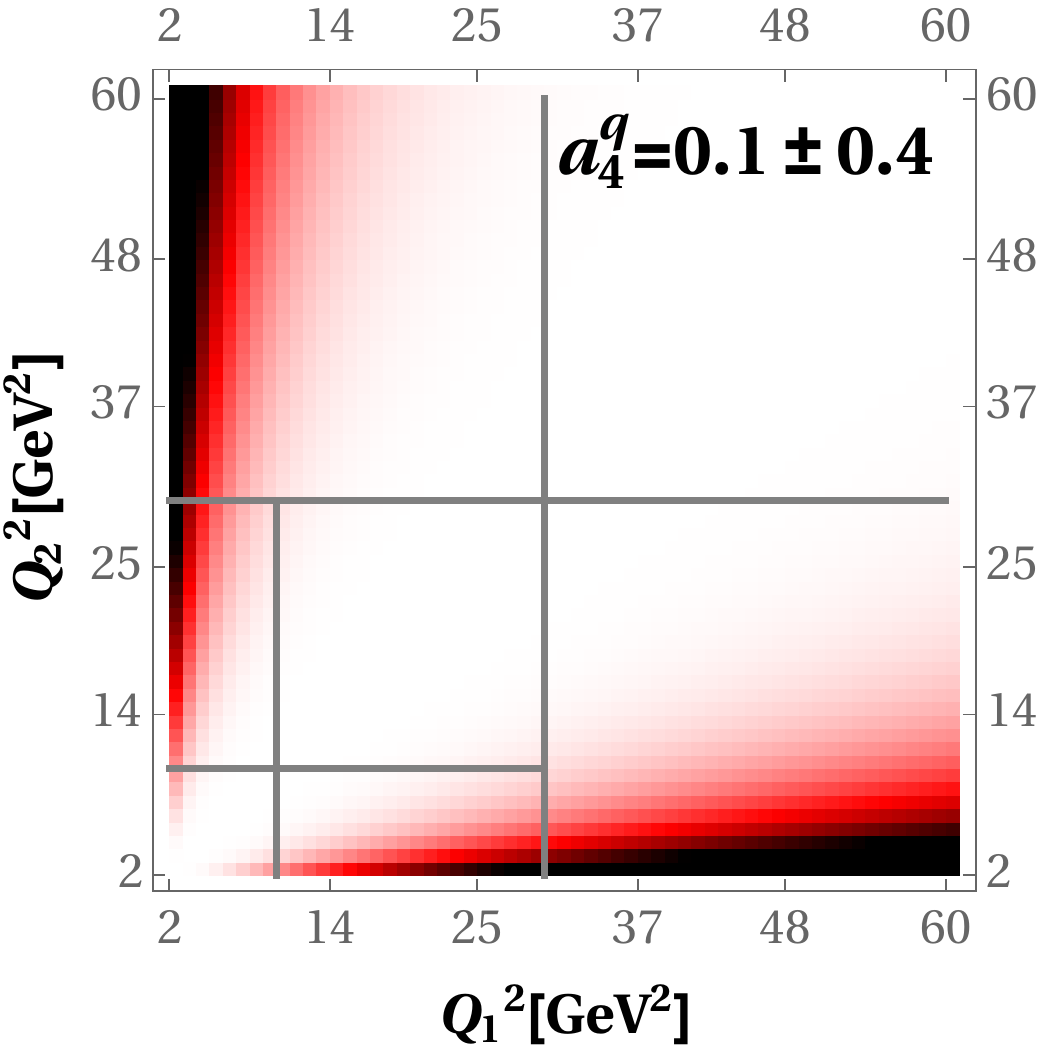}
\includegraphics[width=0.33\textwidth]{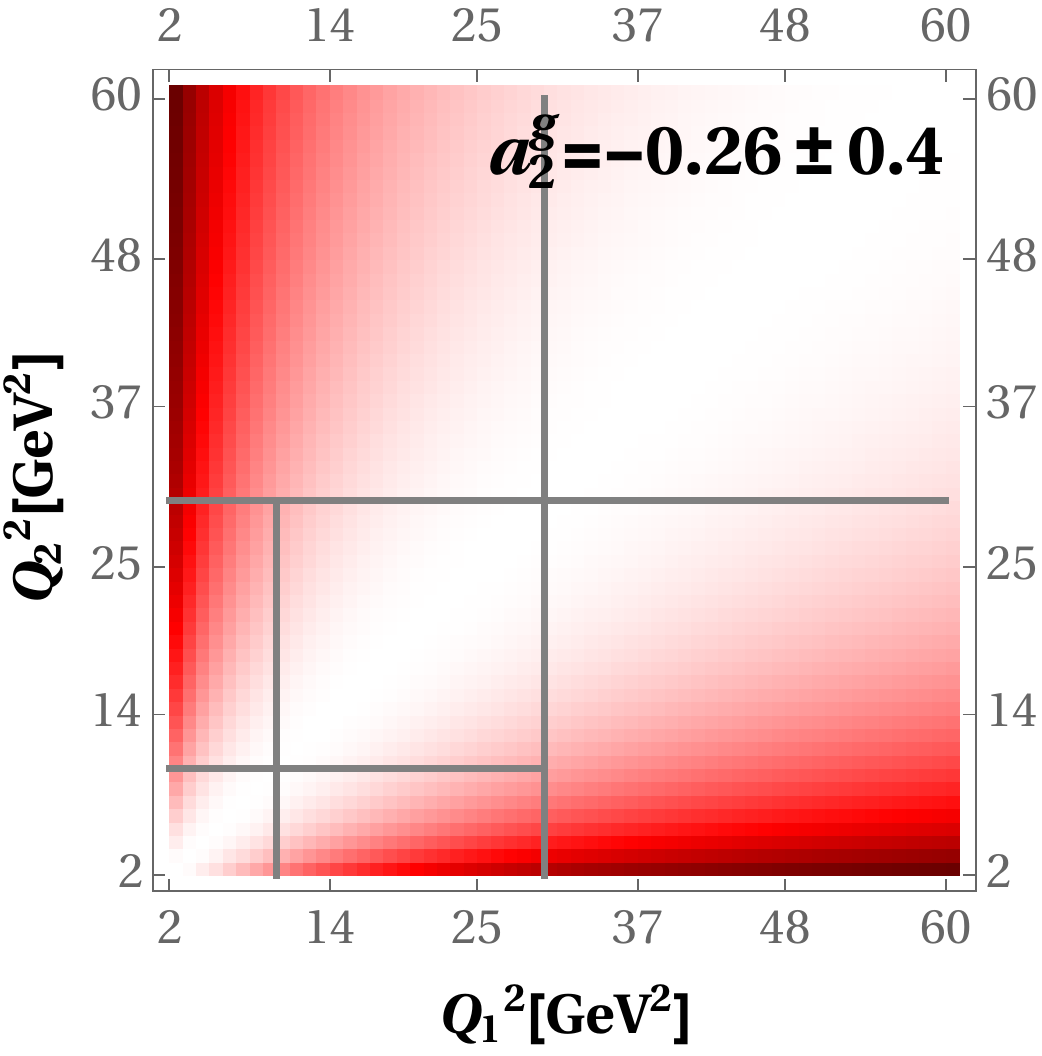}
\caption{The cross-section variation with respect to the change of a parameter in the plane $(Q_1^2,Q_2^2)$. Gray lines show the binning of the data. The values are adjusted to the intensity of the color as in Fig. \ref{fig:h_sensetivity}.}
\label{fig:a_sensetivity} 
\end{figure*}

In this section, we would like to demonstrate the potentials of the double-tag measurements and point out interesting kinematic regions sensitive to one or another physics. In what follows, we use MODEL 1 (with power corrections) with FGC values of state-mixing couplings as the theory input.

\paragraph{Sensitivity to the theory parameters.} First of all, it is interesting to analyze the regions of $Q^2$ regarding their sensitivity to different theory input. With this aim, we vary the values of parameters $a_n$ by a fixed amount $\pm 0.4$, so that $\chi^2/\#\text{points}$ does not significantly deviates from 1, and plot the relative changes of the cross-section (in percentage), see Fig.~\ref{fig:a_sensetivity}. We observe that at the diagonal section $(Q_1^2=Q_2^2)$ the cross-section is practically independent on higher Gegenbauer moments\footnote{In fact, one can check that the convolution of $T_H$ with $n$th Gengebauer moment is proportional to $(Q_1^2-Q_2^2)^{[n/2]}$. Thus, the corrections to asymptotic DA necessarily vanish at the diagonal. }. Their influence on the cross-section increases to the border of the phase-space $Q_i^2\to0$. Naturally, the coefficient $a_2^q$ gives the most important contribution, whereas the contributions of $a_2^g$ and $a_4^q$ are smaller. The dependence on the gluon parameter $a_2^g$ is less rapid than the dependence on the parameter $a_4^q$. Therefore, it influences already the diagonal bins. The measurements of the off-diagonal sector (while staying away from the boundary) would allow one to decorrelate the constants $a_2^g$ and $a_4^q$.

\begin{figure}
  \includegraphics[width=0.33\textwidth]{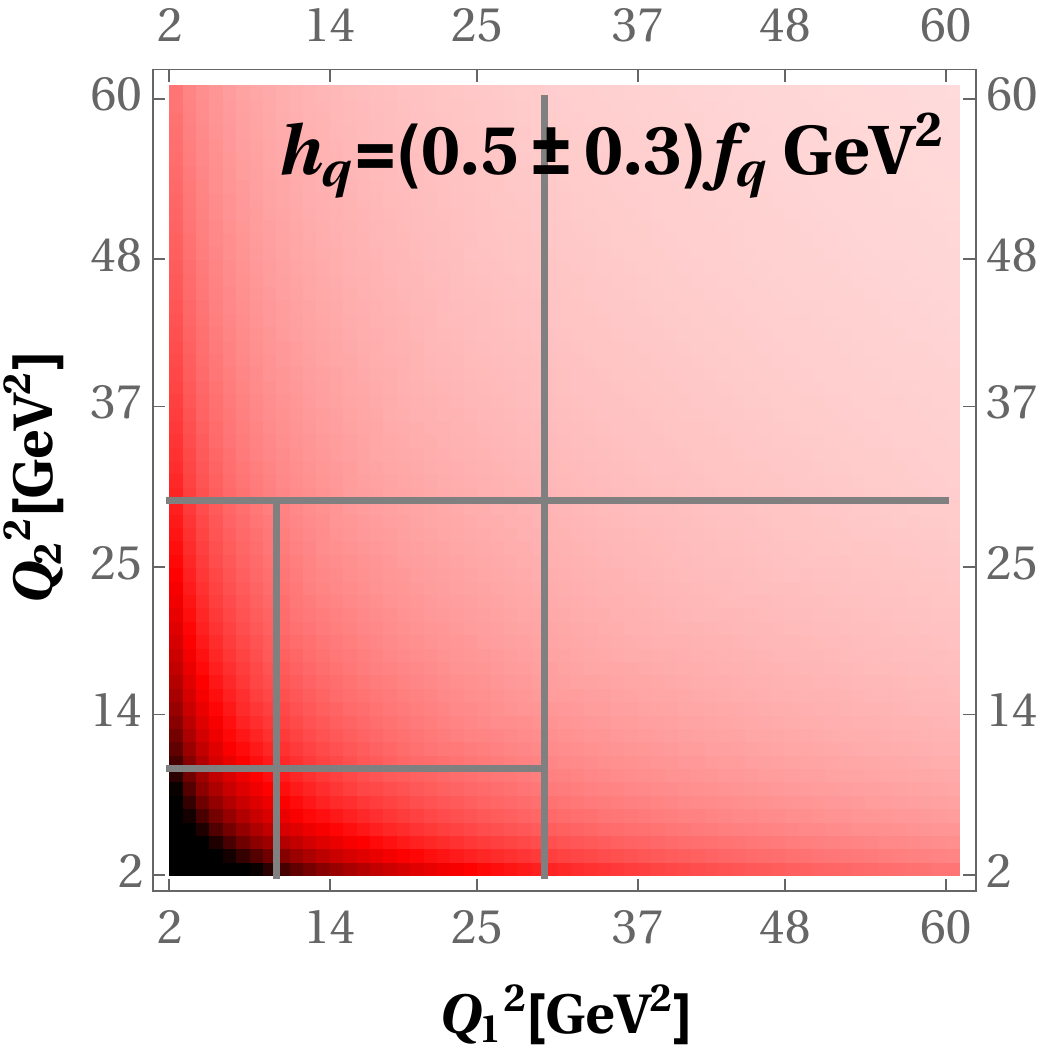}~~~~~
  \includegraphics[width=0.06\textwidth]{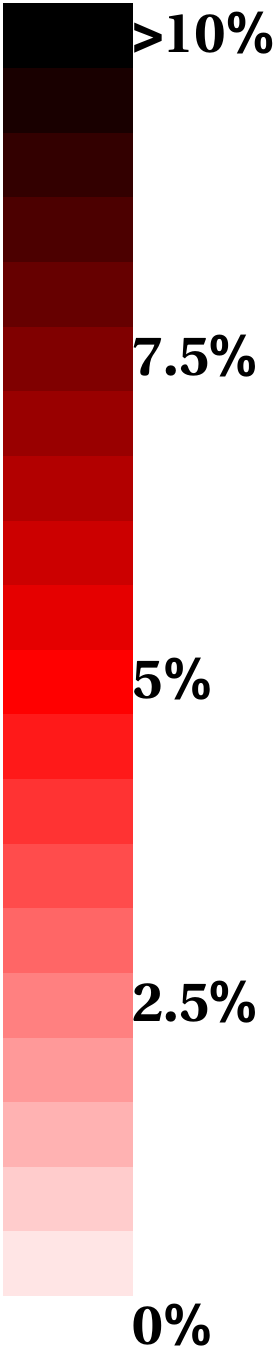}
\caption{The cross-section variation with respect to the change of the twist-3/4 parameter $h_{\eta'}^q$ in the plane $(Q_1^2,Q_2^2)$. Gray lines show the binning of the data. The variation of cross-section by changing  parameter $\delta_{\eta'}^q\pm 0.1 $GeV is practically the same.}
\label{fig:h_sensetivity}
\end{figure}

The similar plot for the sensitivity of the cross-section to the twist-3/4 parameters is shown in Fig.~\ref{fig:h_sensetivity} (Here, we demonstrate only the variation of the parameter $h^q$. The variation of parameter $\delta^q$ results in a almost identical plot). As expected, these parameters are important in the region of small $Q_{1,2}$. What is less expected is that the cross-section's dependence, though small (of the order of $2\%$), still remains at large $Q_{1,2}^2$. 

It is clear that the diagonal values play a special role. In fact, the leading twist contribution of the diagonal bins are entirely determined by the asymptotic quark DA, $\phi^q(x)=6x\bar x$. Thus, \textit{the diagonal bins are the perfect laboratory to determine the couplings $C_{\eta'}^i$ (decay constants)}. Also, by studying the dependence of diagonal values on $Q^2=Q_1^2=Q_2^2$ one can accurately extract the higher-twist parameters, such as $h$ and $\delta$. 

\paragraph{Estimation of parameter error bars.} As we have seen in the previous section, current measurement does not allow meaningful extraction of DA parameters, due to the large error bars and large size of binning at present. Therefore, it is interesting to study the effective size of the error bars with respect to different binning and statistics. To perform this analysis we have generated 100 replicas of pseudo-data and estimated the average errors on the parameter extraction. The result of estimation are presented in Table \ref{tab:errors}.

To generate the pseudo-data we have used the central values predicted by the theory (FGC, MODEL 1), and distributed them with the errors $\alpha\cdot\delta \sigma$, where $\delta \sigma$ is the statistical uncertainty of measurement reported in \cite{BaBar:2018zpn}. The systematic uncertainty is taken to be 12\% (as in \cite{BaBar:2018zpn}). The error estimation is made by averaging over replicas with the boundary of $\chi_{\rm s}^2\pm1$ for a given parameter with $\chi_{\rm s}^2$ equals the number of data points. For $\alpha=1$ the error-estimation produces values similar to one plotted in Fig.~\ref{fig:landscape}, if one ignores the correlation effects. Considering the dynamics of the the error-reduction, we conclude that the original binning is not very efficient. Even reducing statistical uncertainties by factor 10, we are still not able to extract the DA parameters better than an order of magnitude. The reason is that there are only two bins sensitive to variation of these parameters (bins 3 and 4).

We have also considered the pseudo-data generated for an alternative split of the data in 9 bins. The additional energy-bins are shown in Fig.~\ref{fig:energy-bins} by dashed lines. To generate the pseudo-data in this case, we have taken the central values predicted by the theory, and the systematic uncertainty is given by $\alpha \cdot \delta\sigma$, with $\delta\sigma$ taken from the original bin in the percentage (with an original overall systematic uncertainty). With this binning the uncertainties in the extraction of parameters $a_n$ decrease, as it is shown in the second part of the Table \ref{tab:errors}. The uncertainties for parameters $a_4^q$ and $a_2^g$ still remain large. 

In essence, finer binning allows a more accurate determination of the $a_2^q$ constant. It suggests that \textit{with a similar measurement for $\gamma^*\gamma^*\to\eta$, one can put the state-mixing hypothesis for DAs to the test.} Indeed, the diagonal bins would provide accurate determination of the state-mixing constant, whereas, off-diagonal bins determine $a_2^q$ for $\eta$ and $\eta'$ independently.

\begin{table}
\caption{Estimate of the determination uncertainty on the leading twist parameters $a_n$ from the pseudo-data (see text). The parameter $\alpha$ is the relative size of the systematic uncertainty with respect to the original one.}
\label{tab:errors}      
\begin{center}
\begin{tabular}{l||c|c|c}
$\alpha $ & $a_2^q$ & $a_4^q$ & $a_2^g$ \\\hline
\multicolumn{4}{c}{Original binning}
\\\hline
$1$ & $0.1^{+1.02}_{-1.89}$ & $0.1^{+22.5}_{-5.68}$ & $-0.26^{+6.93}_{-13.5}$
\\\hline
$0.75$ & $0.1^{+0.73}_{-1.73}$ & $0.1^{+14.1}_{-5.47}$ & $-0.26^{+4.90}_{-11.4}$
\\\hline
$0.5$ & $0.1^{+0.65}_{-0.97}$ & $0.1^{+3.54}_{-5.89}$ & $-0.26^{+4.40}_{-6.41}$
\\\hline
$0.25$ & $0.1^{+0.56}_{-0.40}$ & $0.1^{+1.88}_{-2.97}$ & $-0.26^{+3.79}_{-2.61}$
\\\hline
$0.1$ & $0.1^{+0.38}_{-0.13}$ & $0.1^{+1.73}_{-0.62}$ & $-0.26^{+2.45}_{-0.82}$
\\\hline
\multicolumn{4}{c}{Extended binning}
\\\hline
$1$ & $0.1^{+0.53}_{-1.89}$ & $0.1^{+2.45}_{-9.33}$ & $-0.26^{+3.38}_{-11.3}$
\\\hline
$0.75$ & $0.1^{+0.39}_{-1.24}$ & $0.1^{+1.79}_{-6.09}$ & $-0.26^{+2.48}_{-7.84}$
\\\hline
$0.5$ & $0.1^{+0.36}_{-0.72}$ & $0.1^{+1.64}_{-3.18}$ & $-0.26^{+2.25}_{-4.43}$
\\\hline
$0.25$ & $0.1^{+0.22}_{-0.32}$ & $0.1^{+0.89}_{-1.22}$ & $-0.26^{+1.47}_{-2.83}$
\\\hline
$0.1$ & $0.1^{+0.12}_{-0.12}$ & $0.1^{+0.48}_{-0.49}$ & $-0.26^{+0.74}_{-0.86}$
\end{tabular}
\end{center}
\end{table}

\section{Conclusion}

We have analyzed the recently measured cross-section of $e^+e^-\to e^+e^-\eta'$ in the double-tag mode. This measurement gives access to the $\eta'$ transition form factor with both non-zero virtualities $F(Q_1^2, Q_2^2)$. It allows one for the first time to test the factorization approach for transition form-factor in the perturbative regime. We have found that the data is in total agreement with the perturbative QCD prediction as well as previous analysis made for form factor with one photon on-shell $F(Q^2,0)$.

Since the provided data have large uncertainties, it is not sufficient for a detailed study of leading twist DA parameters. However, it is sensitive to power corrections (mostly to the meson mass corrections), which should be included in the analysis to describe the data. It also helps in determining the coupling constants and mixing angle in the FKS scheme. In particular, we have shown that values extracted from \cite{Escribano:2005qq} deviate significantly from this measurement. 

We have also presented the study regarding the sensitivity of particular parameters to different regions in the $(Q_1^2,Q_2^2)$ plane. We have demonstrated that the diagonal values ($Q_1^2=Q_2^2$) of the cross-section are practically independent of the higher moments of the leading twist DA, and are entirely described by its asymptotic form. This makes this kinematic region ideal for the determination of the $\eta/\eta'$ decay constants and related parameters. At smaller values of $Q_1^2=Q_2^2$, the diagonal region presents the clean measurement of higher-twist parameters. The sensitivity to higher twist parameters is especially interesting due to the planned accurate extraction of these parameters from QCD lattice calculations \cite{Bali:2018spj}.

The off-diagonal values of the cross-section are important for the determination of parameters of the leading twist DA. We have found that the current binning is not sufficient for such analysis, and in fact, even a decrease of the statistical uncertainty by a factor of 10 could not help in determining these interesting parameters within a reasonable range. The main reason for the large uncertainties is due to the strong correlation between parameters for quark and gluon DAs. One could, however, significantly increase the precision in parameter determination with finer off-diagonal bins. In particular, it is realistic to expect an accurate determination of $a_2^q$ (the second Gegenbauer moment of the leading-twist quark DA). In this case, it would be the first measured parameter for the $\eta'$-meson DA (we recall that nowadays DAs for $\eta$ and $\eta'$ meson are typically taken equal to those of $\pi$-meson, due to a lack of data). Moreover, if the measurement of the form factor for $\gamma^*\gamma^*\to\eta$ becomes available, it will allow us to test the state-mixing hypothesis directly on the level of wave-functions at short distances.

\appendix
\section{Kinematic factors}
\label{app:kinematic}
The function $\Phi$ is originated from the convolution of the photons polarization tensor and the lepton tensor together with the volume of the phase-space integration of an unstable particle. For the process $$e^+(p_a)+e^-(p_b)\to e^+(p_1)+e^-(p_2)+\eta'(p_\eta),$$ it reads
\begin{eqnarray}
\Phi(s,-t_1,-t_2)&=&\frac{1}{\pi}\int dW^2 ds_1ds_2\frac{B}{\sqrt{-\Delta_4}}
\\\nn &&\frac{m_{\eta'}^2}{W}\frac{\Gamma_{\eta'}}{(W^2-m^2_{\eta'})^2+\Gamma^2_{\eta'}m_{\eta'}^2},
\end{eqnarray}
where $s_{1,2}=(p_{1,2}+p_{\eta})^2$, $t_{1,2}=(p_{a,b}-p_{1,2})^2=-Q_{1,2}^2$, $W^2=p^2_{\eta}$, $m_{\eta'}$ and $\Gamma_{\eta'}$ are the mass and decay width of the $\eta'$ state. The factor $B$ has been derived in \cite{Budnev:1974de,Poppe:1986dq} and depends on the angular modulation distribution of electrons. For the integrated case (i.e. for spherical distribution) it reads
\begin{eqnarray}
B&=&\frac{1}{16}\Big(t_1 t_2[(m^2_{\eta'}+4s-2s_1-2s_2+t_1+t_2)^2
\\&&\nn+(t_1+t_2-m^2_{\eta'})^2-4t_1t_2]-4[s(t_1+t_2)
\\\nn &&+(s_2-t_1)(s_1-t_2)-sm^2_{\eta'}]\Big).
\end{eqnarray}
The function $\Delta_4$ is the Gram determinant,
\begin{eqnarray}&&16\Delta_4=
\\\nn &&
\left|
\begin{array}{cccc}
0 & s & -t_1 & s-s_1+t_2 \\
s & 0 & s\!-\!s_2\!+\!t_1 & -t_2 \\
-t_1 & s\!-\!s_2\!+\!t_1 & 0 & \!\!s\!-\!s_1\!-\!s_2\!+\!m_\eta^2 \\
s\!-\!s_1\!+\!t_2 & -t_2 & \!\!s\!-\!s_1\!-\!s_2\!+\!m_\eta^2 & 0
\end{array}
\right|.
\end{eqnarray}
Its null-lines define the boundary of the integration over $s_{1,2}$.

In the narrow-width approximation the integral over $W$ can be removed and the factor simplifies (see also \cite{Druzhinin:2010er})
\begin{eqnarray}
\Phi(s,-t_1,-t_2)&=&\int ds_1ds_2\frac{B}{\sqrt{-\Delta_4}}.
\end{eqnarray}
This integral can be taken explicitly as elementary functions.

\begin{acknowledgements}
We thank V.Braun for multiple discussions, multiple remarks and general enthusiasm. A.V. also thanks V.P.Druzhinin for correspondence.
\end{acknowledgements}

\bibliographystyle{spphys}       
\bibliography{literature}   

\end{document}